\documentclass[12pt,epsf,a4paper
]{article}
\usepackage{graphicx,amsmath,amssymb}
\usepackage[height=25cm,width=16.5cm]{geometry}
\usepackage{bm}
\usepackage{amsfonts}
\usepackage[mathscr]{eucal}
\usepackage{ascmac}
\usepackage{cite}
\usepackage{here}
\usepackage[usenames,dvipsnames]{color}
\usepackage{subfigure}

\usepackage[charter]{mathdesign}

\begin{document}

\def\a{\alpha}
\def\b{\beta}
\def\c{\varepsilon}
\def\d{\delta}
\def\e{\epsilon}
\def\f{\phi}
\def\g{\gamma}
\def\h{\theta}
\def\k{\kappa}
\def\l{\lambda}
\def\m{\mu}
\def\n{\nu}
\def\p{\psi}
\def\q{\partial}
\def\r{\rho}
\def\s{\sigma}
\def\t{\tau}
\def\u{\upsilon}
\def\v{\varphi}
\def\w{\omega}
\def\x{\xi}
\def\y{\eta}
\def\z{\zeta}
\def\D{\Delta}
\def\G{\Gamma}
\def\H{\Theta}
\def\L{\Lambda}
\def\F{\Phi}
\def\P{\Psi}
\def\S{\Sigma}

\def\o{\over}
\def\beq{\begin{eqnarray}}
\def\eeq{\end{eqnarray}}
\newcommand{\gsim}{ \mathop{}_{\textstyle \sim}^{\textstyle >} }
\newcommand{\lsim}{ \mathop{}_{\textstyle \sim}^{\textstyle <} }
\newcommand{\vev}[1]{ \left\langle {#1} \right\rangle }
\newcommand{\bra}[1]{ \langle {#1} | }
\newcommand{\ket}[1]{ | {#1} \rangle }
\newcommand{\EV}{ {\rm eV} }
\newcommand{\KEV}{ {\rm keV} }
\newcommand{\MEV}{ {\rm MeV} }
\newcommand{\GEV}{ {\rm GeV} }
\newcommand{\TEV}{ {\rm TeV} }
\def\diag{\mathop{\rm diag}\nolimits}
\def\Spin{\mathop{\rm Spin}}
\def\SO{\mathop{\rm SO}}
\def\O{\mathop{\rm O}}
\def\SU{\mathop{\rm SU}}
\def\U{\mathop{\rm U}}
\def\Sp{\mathop{\rm Sp}}
\def\SL{\mathop{\rm SL}}
\def\tr{\mathop{\rm tr}}

\def\IJMP{Int.~J.~Mod.~Phys. }
\def\MPL{Mod.~Phys.~Lett. }
\def\NP{Nucl.~Phys. }
\def\PL{Phys.~Lett. }
\def\PR{Phys.~Rev. }
\def\PRL{Phys.~Rev.~Lett. }
\def\PTP{Prog.~Theor.~Phys. }
\def\ZP{Z.~Phys. }

\newcommand{\bea}{\begin{eqnarray}}   
\newcommand{\eea}{\end{eqnarray}}
\newcommand{\bear}{\begin{array}}  
\newcommand {\eear}{\end{array}}
\newcommand{\bef}{\begin{figure}}  
\newcommand {\eef}{\end{figure}}
\newcommand{\bec}{\begin{center}}  
\newcommand {\eec}{\end{center}}
\newcommand{\non}{\nonumber}  
\newcommand {\eqn}[1]{\beq {#1}\eeq}
\newcommand{\la}{\left\langle}  
\newcommand{\ra}{\right\rangle}
\newcommand{\ds}{\displaystyle}
\def\SEC#1{Sec.~\ref{#1}}
\def\FIG#1{Fig.~\ref{#1}}
\def\EQ#1{Eq.~(\ref{#1})}
\def\EQS#1{Eqs.~(\ref{#1})}
\def\TEV#1{10^{#1}{\rm\,TeV}}
\def\GEV#1{10^{#1}{\rm\,GeV}}
\def\MEV#1{10^{#1}{\rm\,MeV}}
\def\KEV#1{10^{#1}{\rm\,keV}}
\def\lrf#1#2{ \left(\frac{#1}{#2}\right)}
\def\lrfp#1#2#3{ \left(\frac{#1}{#2} \right)^{#3}}
\def\REF#1{Ref.~\cite{#1}}
\newcommand{\osc}{{\rm osc}}
\newcommand{\ed}{{\rm end}}
\def\dda#1{\frac{\partial}{\partial a_{#1}}}
\def\ddat#1{\frac{\partial^2}{\partial a_{#1}^2}}
\def\dd#1#2{\frac{\partial #1}{\partial #2}}
\def\ddt#1#2{\frac{\partial^2 #1}{\partial #2^2}}
\def\lrp#1#2{\left( #1 \right)^{#2}}

\def\mi{m_{\phi}}
\def\mpl{M_{\rm pl}}

\makeatletter
\@addtoreset{equation}{section}
\def\theequation{\thesection.\arabic{equation}}
\makeatother
\newcommand{\Slash}[1]{{\ooalign{\hfil/\hfil\crcr$#1$}}}

\renewcommand{\thefootnote}{$\blacklozenge$\arabic{footnote}}

\newcommand{\mtrem}[1]{{ \bf $[[ $ MT: #1$ ]]$}}
\newcommand{\rem}[1]{{\color{red} \bf $[[ $#1$ ]]$}}

\begin{titlepage}

\begin{flushright}
UT-14-31\\
\end{flushright}

\vskip 3.5cm
\begin{center}
{\huge \bf 
Axion Models 
with High Scale Inflation
}
\vskip .75in

Takeo Moroi$^{a,b}$,
Kyohei Mukaida$^{a}$,
Kazunori Nakayama$^{a,b}$ and
Masahiro Takimoto$^{a}$


\vskip .35in
\begin{tabular}{ll}
$^{a}$&\!\! {\em Department of Physics, Faculty of Science, }\\
& {\em University of Tokyo,  Bunkyo-ku, Tokyo 133-0033, Japan}\\[.5em]
$^{b}$ &\!\! {\em Kavli Institute for the Physics and Mathematics of the Universe, }\\
&{\em Todai Institute for Advanced Study,}\\
&{\em University of Tokyo,  Kashiwa, Chiba 277-8583, Japan}
\end{tabular}

\vskip .85in

\abstract{

  We revisit the cosmological aspects of axion models.  In the
  high-scale inflation scenario, the Peccei-Quinn (PQ) symmetry is
  likely to be restored during/after inflation.  If the curvature of
  the PQ scalar potential at the origin is smaller than its vacuum
  expectation value; for instance in a class of SUSY axion models,
  thermal inflation happens before the radial component of the PQ
  scalar (saxion) relaxes into the global minimum of the potential and
  the decay of saxion coherent oscillation would produce too much
  axion dark radiation.  In this paper, we study how to avoid the
  overproduction of axion dark radiation with some concrete examples.
  We show that, by taking account of the finite-temperature
  dissipation effect appropriately, the overproduction constraint can
  be relaxed since the PQ scalar can take part in the thermal plasma
  again even after the PQ phase transition.  We also show that it can
  be further relaxed owing to the late time decay of another heavy
  CP-odd scalar, if it is present.
  
}

\end{center}
\end{titlepage}

\section{Introduction}

The axion $a$ appears in the Peccei-Quinn (PQ) mechanism to solve the
strong CP problem in QCD~\cite{Peccei:1977hh,Kim:1986ax} via the
following interaction
\begin{equation}
  {\cal L}_{\rm PQ} = \frac{\alpha_s}{16\pi F_a} 
  a \epsilon^{\mu\nu\rho\sigma} F_{\mu\nu}^{a} F_{\rho\sigma}^{a},
\end{equation}
where $\alpha_s$ is the QCD fine-structure constant, $F_a$ is the
decay constant of axion, and $F_{\rho\sigma}^{a}$ is the
field-strength tensor of gluon.  The axion predicts various effects on
cosmology~\cite{Kawasaki:2013ae}.  In particular, it is a good
candidate of cold dark matter (CDM) in the
universe~\cite{Preskill:1982cy,Turner:1985si}.  If the PQ symmetry is
broken during inflation, the axion obtains isocurvature
fluctuations~\cite{Axenides:1983hj,Seckel:1985tj,Linde:1985yf} and the
high-scale inflation indicated by the BICEP2~\cite{Ade:2014xna} yields
too large isocurvature fluctuation inconsistent with observations~\cite{Kawasaki:2008sn,Hikage:2012be}.\footnote{ See, however,
  Sec.~\ref{sec:dis} for more on this issue.  }

A more natural situation is that the PQ symmetry is restored during
inflation.  Actually, for example, the PQ field can have a (positive)
Hubble mass through the Planck-suppressed coupling to the inflaton,
which easily stabilizes the PQ scalar at the origin during inflation.
Then there is no axion isocurvature problem.  Instead, the PQ symmetry
breaking after inflation produces topological defects: axionic strings
and axionic domain walls.  The formation of stable domain walls cause
cosmological disaster, hence the domain wall number must be equal to
one so that the string-wall system collapses soon after the
formation~\cite{Sikivie:1982qv}.\footnote{ One can introduce an
  explicit PQ breaking term to make domain walls unstable even in the
  case where the domain wall number is larger than one.  However, it
  changes the potential minimum of the axion and the $\theta$ angle
  reappears.  In order to make domain walls harmless while the
  $\theta$ angle remains small, some level of tuning is
  required~\cite{Hiramatsu:2012sc}. We do not consider such a case in
  this paper.  } It is shown that the axions radiated from the
string-wall system gives a dominant contribution to the relic axion
CDM density and it constrains the PQ scale as~\cite{Hiramatsu:2012gg}
\begin{equation}
  F_a \lesssim (2.0-3.8)\times 10^{10}\,{\rm GeV}.
\end{equation}

If the curvature of the potential of the PQ scalar field at the origin is smaller than its
vacuum expectation value (VEV) times its coupling to particles in
thermal bath, the thermal inflation is likely to take place before the
PQ phase transition.  For instance, in supersymmetry (SUSY), the
radial component of the PQ scalar is often massless in the SUSY limit
and can obtain a soft SUSY breaking mass, which we call saxion.
In a class of SUSY axion models, if the PQ symmetry is restored during
inflation, the saxion is trapped at the origin due to thermal effects
and it causes a brief period of late-time inflation, called thermal
inflation~\cite{Yamamoto:1985rd,Lazarides:1985ja,Lyth:1995hj,
  Choi:1996vz,Chun:2000jr,Kim:2008yu}, before the PQ phase transition.
Even in the non-SUSY case, thermal inflation may take place if the
self coupling constant of the PQ scalar is smaller than its coupling
to particles in thermal bath. (We call the
  radial component of the PQ scalar as saxion even in the non-SUSY
  case.) After thermal inflation ends, the saxion coherent oscillation
dominates the universe and finally the saxion decay reheats the
universe.

However, this scenario does not always work.
This is because the saxion often dominantly decays into the axion pair,
resulting in the axion dominated universe, which contradicts with observational constraint on the amount of dark radiation.

In this paper, we study how to avoid the overproduction of axion dark
radiation with some examples.  We show that, by taking account of the
thermal dissipation effect on the
saxion~\cite{Moroi:2012vu,Mukaida:2012qn,Moroi:2013tea}, interactions
with the thermal plasma can efficiently dissipate the energy of the
saxion coherent oscillation into thermal plasma without producing too
much axionic dark radiation: namely the PQ scalar can participate in
the thermal plasma again even after the PQ phase transition.  In
addition, we also point out that if there exists a heavy CP-odd
scalar, whose existence depends on the stabilization mechanism of the
saxion potential, the overproduction constraint can be further relaxed
owing to its late time decay.

On the other hand, in a class of SUSY axion models, if the PQ scalar
participates in the thermal plasma after the PQ phase transition, the
axino that might be protected by the (approximate) R-parity will be
thermally populated, and it could potentially cause the overproduction
problem depending on the stabilization of saxion.  We show that the
axino overproduction can be avoided with some concrete models.

\section{Basic ingredients}  \label{sec:cos}

\subsection{
Decay and dissipation rates in thermal bath
}

In this paper we consider the hadronic axion model~\cite{Kim:1979if}.
We discuss the saxion dynamics with the following Lagrangian
\begin{equation}
  \mathcal L =|\partial \phi|^2
  - (\lambda \phi \Psi\bar \Psi + {\rm h.c.})
  + m_\phi^2|\phi|^2 - V_{\rm stab}(\phi),
\end{equation}
where $\Psi$ and $\bar\Psi$ are PQ quarks with (anti-)fundamental
representations of color SU(3), and $V_{\rm stab}$ denotes the
potential term that stabilizes the saxion $\phi$ at appropriate scale
$\langle|\phi|\rangle \equiv f_a = F_a/\sqrt{2}$. We take $f_a\sim 10^{9-10}$\,GeV to avoid the
cosmological and astrophysical constraints.  In this section, we
concentrate on rather general aspects of the saxion dynamics which
does not depend on the detail of $V_\text{stab}$.  In the next section
we will discuss $V_{\rm stab}$.

Let us suppose that $\phi$ is placed at the origin during inflation
due, for example, to the Hubble-induced mass term.  Note that even if
$\phi$ is placed far from the origin during inflation, it is
eventually trapped at the origin $\phi = 0$ due to the efficient
particle production at the enhanced symmetry point and its
back reaction onto $\phi$~\cite{Moroi:2013tea} unless the coupling,
$\lambda$, or the reheating temperature, $T_\text{R}$, is suppressed.
(See also Sec.~\ref{sec:dis}.)  Since $\Psi$ and $\bar\Psi$ are massless
there, they are thermalized after the reheating.  It generates the
thermal mass for $\phi$ as $m_T \sim \lambda T$.  Thermal inflation
takes place at the temperature $T_{\rm end} < T < T_{\rm beg}$ where
$T_{\rm beg} \sim \sqrt{m_\phi f_a}$ and $T_{\rm end} \sim
m_\phi/\lambda$.\footnote{Here we have assumed that the reheating
  temperature is higher than $T_{\rm beg}$.}

After thermal inflation ends, the saxion begins a coherent oscillation
around the potential minimum.  The perturbative decay rate of the
saxion into the axion pair is given by~\cite{Chun:1995hc}
\begin{equation}
  \Gamma_{\phi \to 2a} = \frac{1}{64\pi} \frac{m_s^3}{f_a^2},
\end{equation}
where $m_s$ is the saxion mass around the potential minimum, which is
same order of $m_\phi$.  In the following, we do not distinguish $m_s$
from $m_\phi$ unless otherwise stated since it depends on the
stabilization of PQ scalar potential.  It also decays into the gluon
pair, but such a decay mode is subdominant because such a process is
loop-suppressed.  Therefore the universe would be dominated by the
axion dark radiation if the saxion perturbatively decays into the
axion pair.

Fortunately, the axion overproduction constraint may be relaxed by
appropriately taking account of thermal dissipation effect on the
saxion in the thermal
plasma~\cite{Moroi:2012vu,Mukaida:2012qn,Moroi:2013tea}.  The dominant
dissipation comes from the interaction with the thermal plasma via the
dimension five operator suppressed by $f_a$, which is obtained from
integrating out $\Psi$ and $\bar\Psi$ around $\left< |\phi| \right> =
f_a$.  Then, as one can guess from the dimensional analysis, the
reaction rate with the thermal plasma may be proportional to
$T^3/f_a^2$~\cite{Anisimov:2000wx}.  In fact, for $m_\phi \gg g_s^2
T$, the Landau cut contribution also becomes important, which
corresponds to the inverse processes of thermal saxion
production~\cite{Graf:2012hb} proportional to $T^3 / f_a^2$.\footnote{
  For a larger saxion mass, $m_\phi \gg g_s T$, the perturbative decay
  into two gluons becomes important.
} On the opposite limit, the pole contribution regulated by its
thermal width gives the dominant contribution, and the detailed
resummation leads to the following dissipation rate of the saxion
coherent oscillation \cite{Moore:2008ws,Laine:2010cq}:
\begin{equation}
  \Gamma^{\rm (dis)}_\phi \simeq \frac{b\alpha_s^2 T^3}{f_a^2}\times
  \begin{cases}
    1 &\text{for}~~ m_\phi \ll g_s^4 T \\
    \sqrt{g_s^4 T / m_\phi} &\text{for}~~ g_s^4 T \ll m_\phi \ll g_s^2 T
  \end{cases},
\end{equation}
where $b$ is a numerical constant.\footnote{
  Here we have assumed that
  the oscillation period is much slower than the interaction time
  scale of the thermal plasma: $m_s \ll T$,
  which is marginally  satisfied during the course of dynamics.
} In the case of quark gluon plasma, $b\simeq 1/(32\pi^2 \log \alpha_s^{-1})$ \cite{Laine:2010cq}; we
adopt this value of $b$ in our numerical calculation.\footnote
{Strictly speaking, if the PQ quarks are also charged under
  SU(2)$_\text{L}$, there is a contribution from the weak gauge
  interaction.  In addition, in a class of SUSY axion models, there
  exist contributions from PQ squarks and gauginos.  We neglect those
  contributions in the following discussion since they are
    model-dependent.}

The axions, which are produced non-thermally by the decay of the
saxion oscillation, also interact with the thermal plasma.  Since
their typical energy/momentum are given by $m_\phi$ that may be much
smaller than the cosmic temperature at their production, the dissipation rate
of non-thermally produced axions might be different from axion thermal
production rate studied in Refs.~\cite{Graf:2010tv,Salvio:2013iaa}.
For $p \lesssim g_s^2 T$, where $p$ is the typical momentum of the axion, the Landau cut contribution which
corresponds to the axion thermal production may be suppressed.
Rather, the pole contribution regulated by its thermal width may give
the dominant contribution as in the case of the above saxion
oscillation.  However, contrary to the saxion case, the interaction
term can be expressed as $aF_{\mu\nu}^a\tilde F^{a\mu\nu}= a\partial_\mu K^\mu\sim (\partial_\mu a) K^\mu$, and hence the axion
energy/momentum times temperature should be picked up rather than the
thermal mass squared of gauge bosons.  As a result, we expect a
suppression factor, $p^2 / (g_s^4 T^2)$, for the non-thermally
produced axion dissipation rate compared with the saxion dissipation
rate.  Using one-loop computation with the Breit-Wigner
  approximation for the spectral function of gauge bosons, we checked
  that the suppression factor exists.  (See Appendix \ref{sec:app}.)
  Our estimation of the axion dissipation rate is given by
\begin{align}
  \Gamma_a^\text{(dis)} = \frac{\alpha_s^2 T^3}{32 \pi^2 f_a^2}
  \times
  C
  \frac{p^2}{g_s^4 T^2} f(x),
  \label{Gamma_adis}
\end{align}
where $x=p/(g_s^4 T)$, and the function $f(x)$ is
  given in the Appendix [Eq.~\eqref{eq:function}].
In our numerical analysis, we take
\begin{align}
  p = \frac{m_\phi}{2} \text{min} \left( 1, \frac{a_\text{dec}}{a(t)} \right),
\end{align}
with $a_\text{dec}$ and $a(t)$ being the scale factor at the cosmic
time $\Gamma_{\phi\rightarrow 2a}^{-1}$ and at the time $t$,
  respectively.  
In addition, we note here that Eq.\ \eqref{Gamma_adis} merely provides
an order-of-estimate of the dissipation rate of axion.  In order to
take account of the uncertainty, we explicitly introduce the parameter
$C \sim {\cal O} (1)$. 

\subsection{
Boltzmann equations
}

After the thermal inflation, the energy densities of the saxion
oscillation, axion, and radiation, which are denoted as $\rho_\phi$,
$\rho_a$, and $\rho_r$, respectively, evolve as
\begin{align}
  &\dot\rho_\phi + 3H \rho_\phi = 
  - (\Gamma_\phi^{\rm (dis)} + \Gamma_{\phi\to 2a}) \rho_\phi
  \label{rhophidot}
  \\
  &\dot\rho_a + 4H\rho_a = 
  \Gamma_{\phi\to 2a}\rho_\phi - \Gamma_a^{\rm (dis)}\rho_a,
  \label{rhoadot}
  \\
  &\dot\rho_r + 4H\rho_r = 
  \Gamma_\phi^{\rm (dis)}\rho_\phi + \Gamma_a^{\rm (dis)}\rho_a,
  \label{rhordot}
\end{align}
where $H$ is the expansion rate of the universe.

Thus, after a few Hubble times after the PQ phase transition, due to the dissipation effect, the
temperature increases to
\begin{align}
  T_\text{max}
  \sim \min \left[ T_c, \,
    \left( \frac{30}{\pi^2 g_\ast} \right)^{\frac{1}{4}}
    \sqrt{m_\phi f_a}\right],
\end{align}
where
\begin{align}
  T_c = \left( \frac{30}{\pi^2 g_\ast} \right)
  \frac{b\alpha_s^2 m_\phi M_P}{f_a} 
  \sim 10^7\,{\rm GeV} \left( \frac{200}{g_\ast} \right)
  \left( \frac{m_\phi}{1\,{\rm PeV}} \right)
  \left( \frac{10^{10}\,{\rm GeV}}{f_a} \right),
\end{align}
with $M_P\simeq 2.4\times 10^{18}\ {\rm GeV}$ being the reduced Planck
scale.
Notice that $T_\text{max}\sim T_c$
when the dissipation rate is smaller than the expansion rate for
$T\sim T_c$ so that the effect of the dissipation is ineffective.  If
the dissipation rate can become larger than the Hubble parameter, the
energy density of the saxion coherent oscillation is expected to be
transferred to the radiation within a few Hubble time.  In other
words, the PQ scalar takes part in the thermal plasma again.  The
condition is written as
\begin{align}
  \label{eq:cond1}
  \left.\frac{\Gamma^{\rm (dis)}_\phi}{H}\right |_{T=T_{c}} \sim 
  \left( \frac{30}{\pi^2 g_\ast} \right)^3
  \frac{( b\alpha_s^2 m_\phi^{1/2} )^4 M_P^4}{f_a^6}
  \sim  
  \left( \frac{200}{g_\ast} \right)^3
  \left( \frac{m_\phi}{3\,{\rm PeV}} \right)^2
  \left( \frac{10^{10}\,{\rm GeV}}{f_a} \right)^6 \gtrsim 1.
\end{align}

In the above estimation, we have implicitly assumed that the saxion
dominates the universe before the dissipation becomes effective.
However, in general, it is possible that the saxion dominantly decays
into axions at first, and then the axions which dominate the universe
transport their energy into radiation.  The decay mode into axion
dominates when
\begin{align}
  \left. \frac{\Gamma^\text{(dis)}_\phi}{\Gamma_{\phi \to 2a}} 
  \right|_{T \sim \sqrt{m_\phi f_a}}
  &= \left( \frac{30}{\pi^2 g_\ast} \right)^{3/4}
  \left( 64\pi b \alpha_s^2 \right) 
  \left( \frac{f_a}{m_\phi} \right)^{3/2} \nonumber \\
  \label{eq:axion_dom}
  &\sim 10
  \left( \frac{200}{g_\ast} \right)^{4/3}
  \left( \frac{10\,\text{PeV}}{m_\phi} \right)^{3/2}
  \left( \frac{f_a}{10^{10}\,\text{GeV}} \right)^{3/2}
  \lesssim 1.
\end{align}
Note that the ratio between the saxion decay rate into the axion pair and
  the Hubble parameter $H$ at the end of thermal inflation is given by
\begin{equation}
  \left.\frac{\Gamma_{\phi \to 2a}}{H}\right |_{T=T_{\rm end}} 
  \sim \frac{m_\phi^2 M_P}{64\pi f_a^3} \sim 
  10^{-2} \left( \frac{m_\phi}{1\,{\rm PeV}} \right)^2
  \left( \frac{10^{10}\,{\rm GeV}}{f_a} \right)^3.
\end{equation}
If this ratio is larger than one, and also if $\Gamma_{\phi
  \to 2a}$ is larger than $\Gamma_\phi^{\rm (dis)}$, the saxion soon
decays into the axion pair within one Hubble time.
In this case, the temperature can be estimated as
\begin{align}
  T_\text{max} \sim \min\left[ \delta T_{c},\, 
    \left( \frac{30}{\pi^2 g_\ast} \right)^{\frac{1}{4}}
    \sqrt{m_\phi f_a} 
  \right],
\end{align}
where $\delta$ is a numerical factor; $\delta \propto
(f_a /M_P)^{n/(n+1)}$ with $n = 0$ -- $2$ for $f(x) \propto
  x^{n-2}$.  The condition for axion thermalization is given by
\begin{align}
 \label{eq:cond2}
  \left.\frac{\Gamma^{\rm (dis)}_a}{H}\right |_{T=\delta T_{c}} \sim 
  \left( \frac{30}{\pi^2 g_\ast} \right)^3
  \frac{( \delta b\alpha_s^2 m_\phi^{1/2} )^4 M_P^4}{f_a^6}
  \sim  
  \delta^4
  \left( \frac{200}{g_\ast} \right)^3
  \left( \frac{m_\phi}{3\,{\rm PeV}} \right)^2
  \left( \frac{10^{10}\,{\rm GeV}}{f_a} \right)^6 \gtrsim 1.
\end{align}

If these conditions [Eqs.~\eqref{eq:cond1}, \eqref{eq:cond2}] are satisfied, the temperature increases up to 
$T\sim T_\text{max}$ within a few Hubble time after thermal inflation and the
saxion coherent oscillation disappears without producing too much
axion dark radiation.
As one can see, in the case of the axion CDM,
the mass scale $m_\phi$ should be larger than ${\cal O}(1)$\,PeV
in order for the PQ scalar to take part in the thermal plasma,\footnote{
\label{fn:thermal}
Thermal saxions decay into axions immediately
when they become non-relativistic,
which results in tiny contribution to dark radiation.
Thermal axions also have tiny contribution to dark radiation.
} which implies the maximum temperature to be $T_\text{max} \sim {\cal
  O}(10^{7})$\,GeV for $m_\phi \sim {\cal O} (1)$\,PeV.  
 Note that such a value of $m_\phi$ is suggested by high-scale SUSY breaking models~\cite{Ibe:2011aa,ArkaniHamed:2012gw,Arvanitaki:2012ps}.
The detailed discussion on which parameters the overproduction of axion dark
radiation can be avoided is performed with some examples in the
following, since it is model dependent.

Before going into details of models, let us briefly give general
comments.  PQ quarks, $\Psi$ and $\bar\Psi$, become massive after the
saxion obtains a VEV and they can soon decay if there is a mixing
between PQ quarks and SM fermions~\cite{Moroi:2013tea}.  In a class of
SUSY axion models, for $(m_\phi, f_a) \sim ({\cal O} (1)\,\text{PeV} ,
{\cal O} (10^{10})\,\text{GeV})$, gravitinos are thermally
produced~\cite{Bolz:2000fu}, but its abundance is not so large and it
does not cause cosmological problems for $m_{3/2} \sim \mathcal
O(100)\,$TeV.  The axion multiplet, saxions, axions and axinos, will
also be thermally populated if the dissipative effect is efficient.
In particular, the axino whose stability might be protected by the
(approximate) R-parity must decay well before 
Big-Bang nucleosynthesis (BBN) to avoid
cosmological constraints.  The axino mass crucially depends on the
saxion stabilization model.  Below we will discuss it with two models
(Model 2 and 3).

\section{Cosmology: model-dependent discussion
}

In the following subsections, we study in detail the dynamics of the
saxion after the thermal inflation, and show how the overproduction
constraint of axion dark radiation is relaxed with some bench mark
models.  First, we consider the simple non-SUSY case (Model 1), where
it is shown that the overproduction constraint is relaxed by
interactions with the thermal plasma.  Then, we consider SUSY axion
models with two stabilization mechanisms of the saxion potential:
radiative stabilization (Model 2) and higher dimensional
superpotential with another PQ field (Model 3).  In particular, in the
latter case (Model 3), we show that the overproduction constraint is
further relaxed by the late time decay of the heavy CP-odd scalar.

In addition,
in a class of SUSY axion models,
there exists the axino which might be protected by the (approximate) R-parity,
and hence it could potentially overclose the universe due to the thermal production,
if the dissipation of the PQ scalar is efficient.
Since the axino overproduction problem crucially depends on the stabilization mechanism of the saxion,
we also discuss how to avoid this problem separately with two cases (Model 2 and 3).

\subsection{Model 1}
Let us first consider the non-SUSY case where the saxion potential is
simply given by
\begin{align}
\label{eq:nonsusy}
	V = \frac{m_\phi^2}{2 f_a^2} \left( |\phi|^2 - f_a^2 \right)^2.
\end{align}
In this case, all the necessary ingredients are already given in the
previous subsection.  Hence one can calculate the effective neutrino
number of axion dark radiation for each parameter $(m_\phi, f_a)$ by
numerically solving Eqs.~\eqref{rhophidot} -- \eqref{rhordot}.

\begin{figure}[th]
\centering
\subfigure[{\bf Left Panel}: $(m_\phi, f_a)=(50\, \text{PeV},10^{10}\, \text{GeV})$]{
\includegraphics[width=0.45\columnwidth,clip]{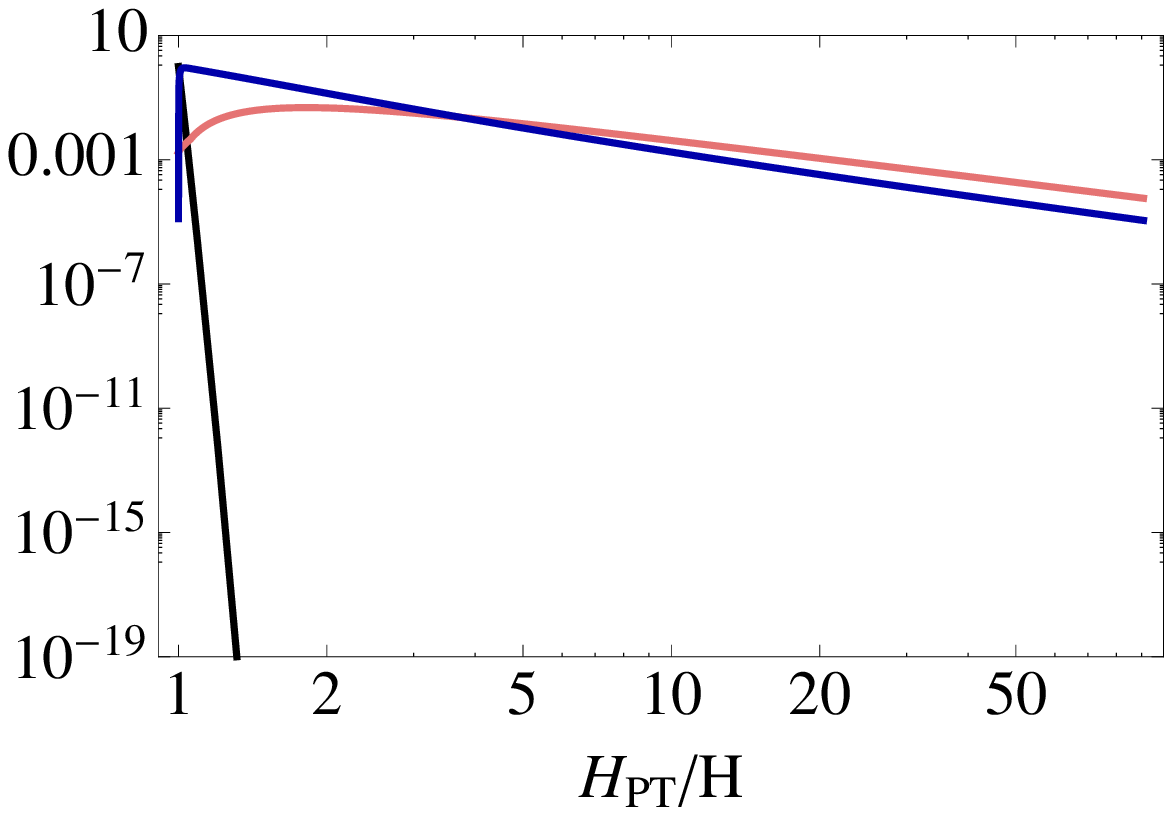}
\label{fig:left}}
\subfigure[{\bf Right Panel}: $(m_\phi, f_a)=(10\, \text{TeV},5\times10^{8}\, \text{GeV})$]{
\includegraphics[width=0.45\columnwidth,clip]{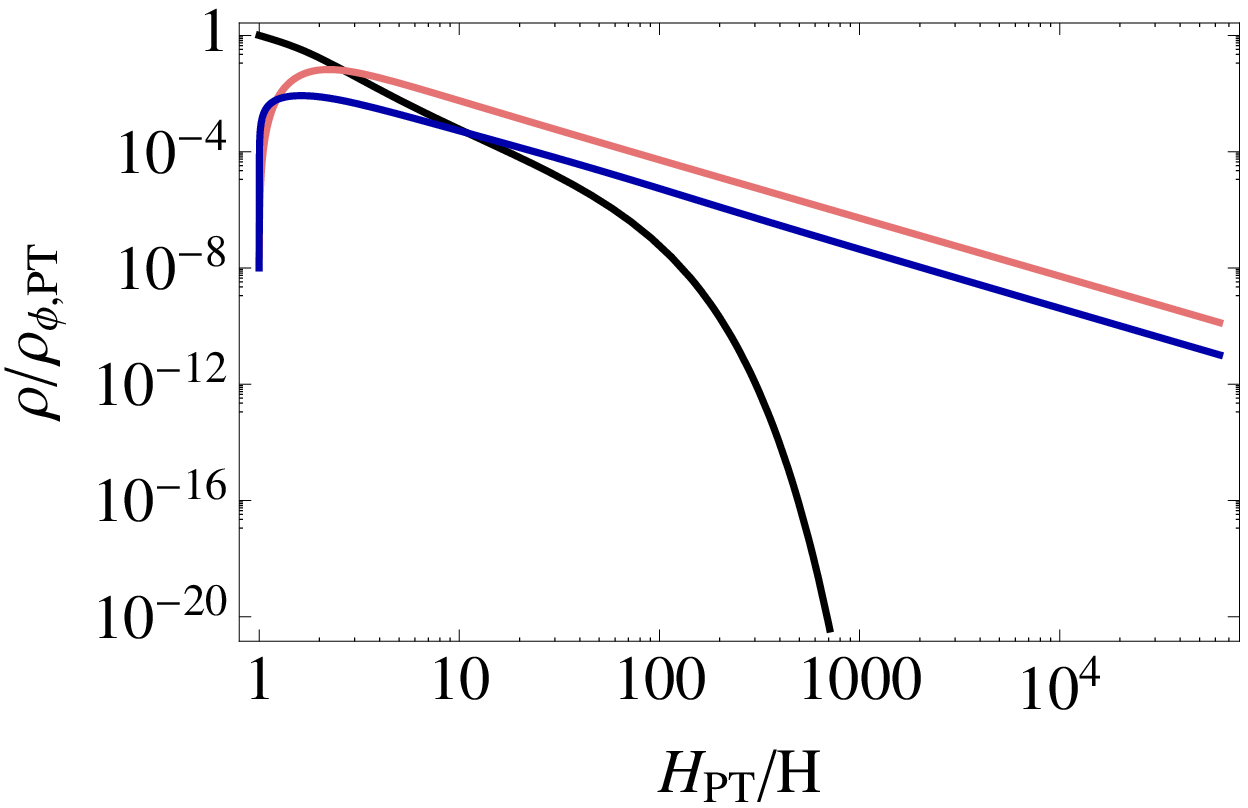}
\label{fig:right}}
\caption{\small 
{\bf Left/Right Panel} shows evolution of various quantities
as a function of $H$ normalized by
one at the PQ phase transition, $H_\text{PT}$:
$\rho_\phi/\rho_{\phi, \text{PT}}$ [black],
$\rho_\text{rad}/\rho_{\phi, \text{PT}}$ [{\color[rgb]{1.000000,0.400000,0.400000}pink}],
$\rho_a/\rho_{\phi,\text{PT}}$ [{\color{blue}blue}];
with $\rho_\bullet$ being the energy density of $\bullet$.
Note that $\rho_\phi$ and $\rho_a$ denote the energy density of the
saxion coherent oscillation and that of
the axion produced from it respectively,
and both does not contain that from thermal plasma.
See also footnote~\ref{fn:thermal}.
}
\label{fig:ev}
\end{figure}

The results are shown in Fig.~\ref{fig:ev}, which shows evolution of
various quantities after the PQ phase transition as a function of
Hubble time $H^{-1}$ normalized by that at the PQ phase transition,
$H_\text{PT}^{-1}$.  The black, pink and blue lines represent
evolution of the energy densities of the saxion coherent oscillation,
radiation, and the non-thermal axion produced from the saxion coherent
oscillation normalized by that of the initial saxion coherent
oscillation, $\rho_{\phi, \text{PT}}$. In the case with $(m_\phi,f_a)
= (50 \,\text{PeV}, 10^{10} \, \text{GeV})$ [Left Panel of
Fig.~\ref{fig:left}], the saxion coherent oscillation first loses its
energy into axion and radiation within a few Hubble times.  Then, the
non-thermally produced axions interact with radiation and reduce their
number.  In this specific parameter, the predicted effective number of
extra radiation is $\Delta N_\text{eff} \sim 0.4$.  On the other hand,
in the case with $(m_\phi,f_a) = (10 \,\text{TeV}, 5 \times 10^{8} \,
\text{GeV})$ [Right Panel of Fig.~\ref{fig:right}], the saxion
coherent oscillation can lose almost all the energy before the decay
into axions dominates.
The resultant axion dark radiation is $\Delta N_\text{eff} \sim 0.2$.

\begin{figure}[th]
\centering
\subfigure[{\bf Left Panel}: $T_{\rm end} = m_\phi$]{
\includegraphics[width=0.45\columnwidth,clip]{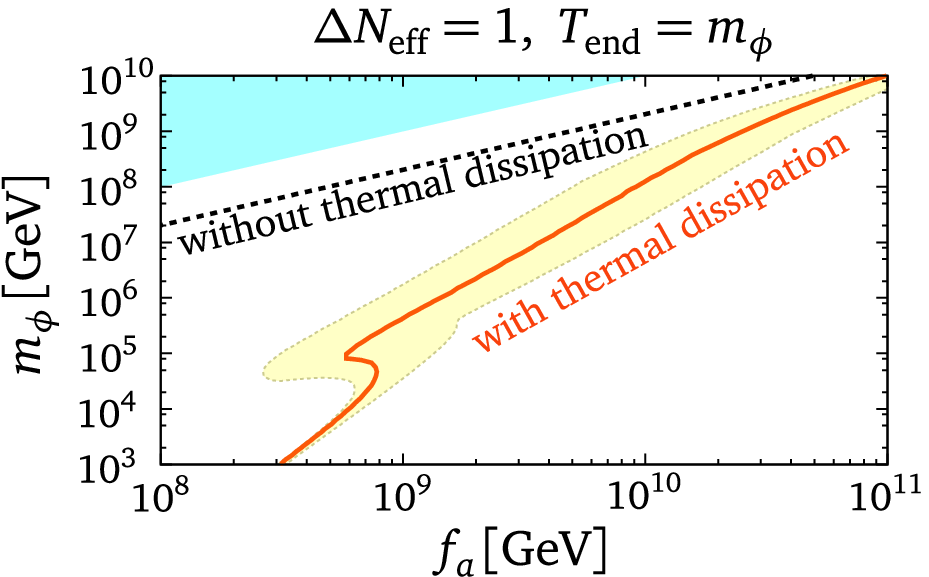}
\label{fig:cntr_l}}
\subfigure[{\bf Right Panel}: $T_{\rm end} = 10m_\phi$]{
\includegraphics[width=0.45\columnwidth,clip]{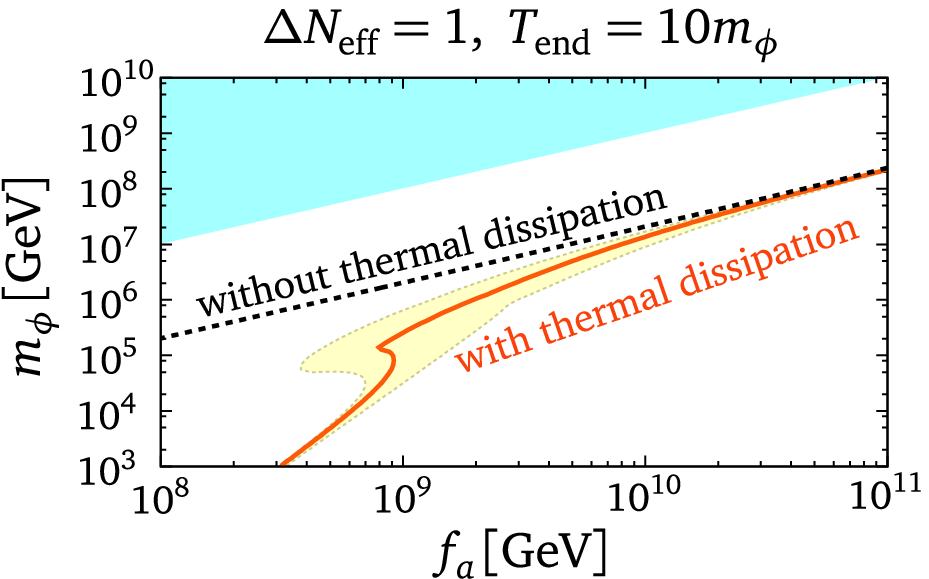}
\label{fig:cntr_r}}
\caption{\small {\bf Left} and {\bf Right Panel} shows a contour of
  $\Delta N_\text{eff} = 1$ as a function of $(m_\phi, f_a)$ for
  $T_{\rm end} = m_\phi$ and $T_{\rm end} = 10m_\phi$ respectively.
  The upper and lower boundaries of the band [{\color{Dandelion}yellow}] correspond
  to $C=1$ and $C=10$, respectively.  Above the band, $\Delta
  N_\text{eff}$ becomes smaller than one.  The black dashed line
  represents the contour $\Delta N_\text{eff} = 1$ without the thermal
  dissipation.  In the {\color{cyan}blue} shaded region, $T_{\rm end}$ becomes
    (typically) larger than the postulated value due to the self
    coupling of $\phi$, and hence such a region is irrelevant.}
\label{fig:cntr}
\end{figure}

Fig.~\ref{fig:cntr} shows contour of $\Delta N_{\rm eff}=1$ on
$(m_\phi, f_a)$ plane for $T_{\rm end} = m_\phi$ (corresponding to
$\lambda \sim 1$) (left panel) and $T_{\rm end} = 10m_\phi$
(corresponding to $\lambda \sim 0.1$) (right panel).\footnote
{If $f_a\lesssim T_{\rm end}$, $T_{\rm end}$ becomes larger than
    the postulated value due to the thermal mass of $\phi$ generated
    by its self coupling.  We do not consider such a parameter region.
    (In Fig.~\ref{fig:cntr}, such a region is shaded in blue.)}
In the figure, we take account of the fact that our estimation of the
dissipation has uncertainty factor owing to both the model dependence
({\it e.g.,} contributions from other gauge group, gauginos) and
theoretical uncertainties.  Thus, we vary the dissipation rate and
show how much the bound changes; for this purpose, we have varied the
$C$-parameter in Eq.\ \eqref{Gamma_adis}.  The upper and lower
boundaries of the band shown in Fig.\ \ref{fig:cntr} correspond to
$C=1$ and $C=10$, respectively.  Above the band, $\Delta N_\text{eff}$
becomes smaller than one.  The black dashed line represents the
contour $\Delta N_\text{eff} = 1$ without the thermal dissipation.  We
can see that the region with small enough $\Delta N_{\rm eff}$ is
significantly enlarged by taking account of the effect of axion
dissipation.  In particular, in the case where $\phi$ couples to the
PQ fermions relatively strongly (i.e, $\lambda\sim 1$), $m_\phi$ is
required to be close to $f_a$ if the effect of thermal dissipation is
neglected; this is in order to avoid the saxion domination after the
PQ phase transition.  With the proper inclusion of the effects of
thermal dissipation, $m_\phi$ much smaller than $f_a$ becomes allowed
even if $\lambda\sim 1$.

\subsection{Model 2}

Next, let us consider the SUSY axion model.  First, we focus on the
possibility that the running of the soft mass of $\phi$ induces the
radiative symmetry breaking at an appropriate
scale~\cite{Abe:2001cg,Nakamura:2008ey}.  We consider the following
superpotential:
\begin{equation}
	W_{\rm PQ} = \lambda_Q \phi Q\bar Q + \lambda_L \phi L \bar L, 
\end{equation}
where $\phi$ denotes the PQ scalar, $\bar Q (Q)$ and $L (\bar L)$ are
chiral multiplets in the (anti-)fundamental representation of SU(5).
Then, the renormalization group equations (RGEs) for the Yukawa
coupling constants are
\begin{equation}
\label{eq:rge_1}
\begin{split}
  \frac{d\lambda_Q}{dt} &= \frac{\lambda_Q}{16\pi^2}
  (\gamma_\phi + \gamma_Q + \gamma_{\bar Q}), \\
  \frac{d\lambda_L}{dt} &= \frac{\lambda_L}{16\pi^2}
  (\gamma_\phi + \gamma_L + \gamma_{\bar L}), 
\end{split}
\end{equation}
where $t=\log E$ (with $E$ being the energy scale) and
\begin{equation}
\label{eq:rge_2}
\begin{split}
  \gamma_\phi &= 3\lambda_Q^2 + 2\lambda_L^2,\\
  \gamma_Q = \gamma_{\bar Q} &= \lambda_Q^2
  -\frac{8}{3}g_3^2 -\frac{2}{15}g_1^2 ,\\
  \gamma_L = \gamma_{\bar L}  &= \lambda_L^2 
  -\frac{3}{2}g_2^2 -\frac{3}{10}g_1^2.
\end{split}
\end{equation}
In addition, the renormalization group equations for the soft
SUSY breaking parameters are given by
\begin{equation}
\label{eq:rge_3}
\begin{split}
  \frac{dm_\phi^2}{dt} &= \frac{1}{8\pi^2}\left[
    3\lambda_Q^2(m_\phi^2+m_Q^2+m_{\bar Q}^2) +
    2\lambda_L^2(m_\phi^2+m_L^2+m_{\bar L}^2)
  \right] ,\\
  \frac{dm_Q^2}{dt} = \frac{dm_{\bar Q}^2}{dt} &= \frac{1}{8\pi^2}\left[
    \lambda_Q^2(m_\phi^2+m_Q^2+m_{\bar Q}^2)
  \right] ,\\
  \frac{dm_L^2}{dt} = \frac{dm_{\bar L}^2}{dt} &= \frac{1}{8\pi^2}\left[
    \lambda_L^2(m_\phi^2+m_L^2+m_{\bar L}^2)
  \right],
\end{split}
\end{equation}
where $m_X$ (with $X=Q,\bar{Q},L,\bar{L}$) denote the soft SUSY
breaking scalar mass parameters for the scalar components in
corresponding chiral multiplet.  Here we have ignored gaugino masses
by assuming that gaugino masses are much smaller than those of
sfermions as in the pure gravity mediation model.

\begin{figure}
\centering
\includegraphics[
width=8cm,
clip]{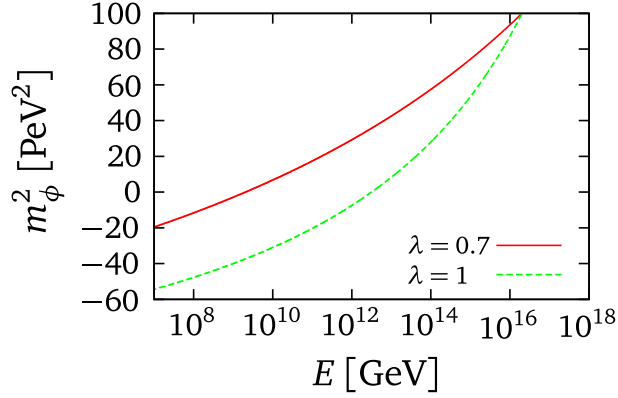}
\caption {\small RGE evolution of the soft mass of the PQ scalar $m_\phi^2$.
  Red-solid and green-dashed lines are for $\lambda_{Q,L}(M_{\rm
    GUT})=0.7$ and $1$, respectively.}
\label{fig:mphi}
\end{figure}

In this model, the dynamics of the PQ scalar after the PQ phase
transition is almost the same as the model 1.  Therefore, from
Fig.~\ref{fig:cntr}, we can understand the parameter region on
$(m_\phi, f_a)$ plane, where the overproduction of axion dark
radiation is avoided.  For instance, the axion CDM implies that the
mass scale, $m_\phi$, should be larger than ${\cal O} (1)$\,PeV.  To
see if the radiative symmetry breaking really occurs,
we have solved the
RGEs with the boundary condition at the GUT scale as $m_\phi^2 = m_Q^2
= m_L^2 = (10\,{\rm PeV})^2$ and $\lambda_Q=\lambda_L= 1.0$ and $0.7$.
The result is shown in Fig.~\ref{fig:mphi}. It is seen that
$m_\phi^2$ becomes negative at some intermediate scale.  In
particular, for $\lambda_{Q,L}\simeq 0.7$, the VEV of $\phi$ can be
around $10^{10}$\,GeV.  Thus the cosmological scenario studied in the
previous section works.

However, this model suffers from the axino overproduction.  In this
model, the axino mass arises only radiatively and is suppressed
compared with the gravitino mass~\cite{Goto:1991gq,Nakamura:2008ey}.  It is
estimated as $m_{\tilde a} \sim (5/16\pi^2)^2\lambda^4 m_{3/2}$.  Thus
it is lighter than the gauginos and it likely becomes the lightest
SUSY particle (LSP).  If the thermal dissipation is effective, the
axion multiplet takes part in the thermal plasma and hence axinos are
thermally populated, which results in overproduction of the axino.
Hence the presence of stable axino LSP is problematic.  Below we list
some possibilities to avoid the axino overproduction.

\begin{itemize}

\item 
If one introduces another LSP, the problem may be avoided.
For example in a singlet extensions of the minimal SUSY standard model~\cite{Ellwanger:2009dp},
the singlino-like neutralino can be the LSP.
In this case the axino can decay into the singlino LSP through the singlino-gaugino mixing, which is suppressed by the SUSY breaking scale. 
The axino decay temperature is then estimated as
\begin{align}
	T_{\tilde a} \sim 10^{-2}\,{\rm GeV} ~
	\lambda^6\left(\frac{(1{\rm TeV})^2}{\mu M}\right)
	\left( \frac{m_{3/2}}{1\,{\rm PeV}} \right)^{3/2} \left( \frac{10^{10}\,{\rm GeV}}{f_a} \right),
\end{align}
where $M$ denotes gaugino mass scale and $\mu$ is the higgsino mass.
In order for the axino to decay well before BBN, relatively light
neutralino mass spectrum is needed.  In addition, singlino also has to
decay before BBN, which requires (small) R-parity violation.  Note
that if the axino decays after the QCD phase transition, the axion
abundance is diluted and upper bound on the PQ scale is
relaxed~\cite{Kawasaki:1995vt}.

\item
The axino may become heavier if there is a sizable $A$-term,
\begin{equation}
	V \supset A_\phi \lambda \phi \Psi \bar\Psi + {\rm h.c.},
\end{equation}
which yields the axino mass as $m_{\tilde a} \sim (\lambda^2/16\pi^2)
A_\phi$~\cite{Goto:1991gq} and it can exceed the anomaly mediation
contribution to the gaugino masses~\cite{Randall:1998uk} if $A_\phi
\sim m_{3/2}$.  Actually we naturally have $A_\phi \sim m_{3/2}$ if
the SUSY breaking field $z$ has no gauge quantum number, as in the
Polonyi model.  The gauginos are also expected to obtain masses of
$\sim m_{3/2}$, but the axino can be heavier if the coupling between
gaugino and Polonyi field happens to be relatively small.  Then the
axino decays into gaugino.  Note that, in this case, the assumption on
RGEs, Eqs.~\eqref{eq:rge_1}--\eqref{eq:rge_3}, is marginally satisfied
since the gauginos are lighter than the gravitino. If the Wino is
LSP, Winos produced by the axino decay annihilate efficiently and its
abundance can become smaller than the DM
abundance~\cite{Moroi:2013sla}.  In this case, however, there arises a
cosmological Polonyi problem~\cite{Coughlan:1983ci,Banks:1993en} which
cannot be solved even in the presence of thermal inflation.  In
Sec.~\ref{sec:dil}, we will discuss this issue and how to solve it.

\end{itemize}

\subsection{Model 3}

Let us consider the following model with two PQ scalar $\phi$ and $\bar\phi$~\cite{Murayama:1992dj}:
\begin{equation}
	W_{\rm PQ} =\lambda \phi \Psi\bar \Psi + \frac{\phi^n \bar\phi}{M^{n-2}}, 
\end{equation}
where $\phi$ and $\bar\phi$ are assumed to have PQ charges of $+1$ and $-n$, respectively, and $M$ is a cutoff scale.
Then the scalar potential is 
\begin{equation}
	V = -m_\phi^2|\phi|^2 + m_{\bar\phi}^2|\bar\phi|^2 + \frac{|\phi|^{2(n-1)}}{M^{2(n-2)}} \left( |\phi|^2 + n^2|\bar\phi|^2  \right)
	+\left\{ (n-2)m_{3/2} \frac{\phi^n\bar\phi}{M^{n-2}} +{\rm h.c.} \right\},
\end{equation}
where the parameters $m_\phi^2$ and $m_{\bar\phi}^2$ are both taken to
be positive.  For computational simplicity, we assume $m_{\bar\phi}^2
\ll m_\phi^2$.  Then the potential minimum lies at
\begin{equation}
\begin{split}
	v_\phi \equiv \langle|\phi |\rangle &\simeq \left[ 
		\left( m_\phi^2 + \frac{(n - 2)^2}{n^2}m_{3/2}^2 \right) \frac{M^{2n-4}}{n}
	\right]^{\frac{1}{2n-2}}, \\
	v_{\bar\phi} \equiv \langle|\bar\phi |\rangle &\simeq  v_\phi \frac{n - 2}{n^{3/2}}
	\left( \frac{m_\phi^2}{m_{3/2}^2} + \frac{(n - 2)^2}{n^2} \right)^{-1/2}.
\end{split}
\end{equation}
Thus for $n=3$, we obtain $v_\phi \sim v_{\bar\phi} \sim \sqrt{m_\phi M} \sim 10^{10}\,$GeV for 
$M \sim 10^{15}\,$GeV and $m_\phi \sim 100$\,TeV.

In this model, the fermionic component of $\phi$ and $\bar\phi$, which
we denote by $\tilde a$ and $\tilde {\bar a}$ and call them as axino,
obtain both Dirac and Majorana masses after PQ scalars get VEVs.  Then
the axino masses are generated as $m_{\tilde a} \sim m_\phi$.  Since
they are as heavy as the gravitino, they decay into the gluino and
gluon:
\begin{equation}
  \Gamma(\tilde a \to \tilde g + g) = 
  \frac{\alpha_s^2}{16\pi^3}
  \frac{m_{\tilde a}^3}{f_a^2}
  \left(1- \frac{m_{\tilde g}^2}{m_{\tilde a}^2} \right)^3.
\end{equation}
The axino decay temperature is then estimated as
\begin{equation}
  T_{\tilde a} \sim 3 \times 10^5\,{\rm GeV}
  \left( \frac{m_{\tilde a}}{1\,{\rm PeV}} \right)^{3/2} 
  \left( \frac{10^{10}\,{\rm GeV}}{f_a} \right).
\end{equation}
Thus,  they are harmless.

Note that the massless axion is a mixture of the angular component of
$\phi$ and $\bar\phi$.  The real components of $\phi$ and $\bar\phi$
are also mixed with each other and both $\phi$ and $\bar\phi$ begins a
coherent oscillation after thermal inflation
ends~\cite{Nakayama:2012zc}.  Although the dissipation effect acts
only on $\phi$, the time scale of the mixing $(\sim m_\phi^{-1})$ is
much faster than that of the cosmic expansion and hence both the
energy density of $\phi$ and $\bar\phi$ are efficiently dissipated.

\begin{figure}[t]
\centering
\subfigure[{\bf Left Panel}: $(m_\phi, f_a)=(50\, \text{PeV},10^{10}\, \text{GeV})$]{
\includegraphics[width=0.45\columnwidth,clip]{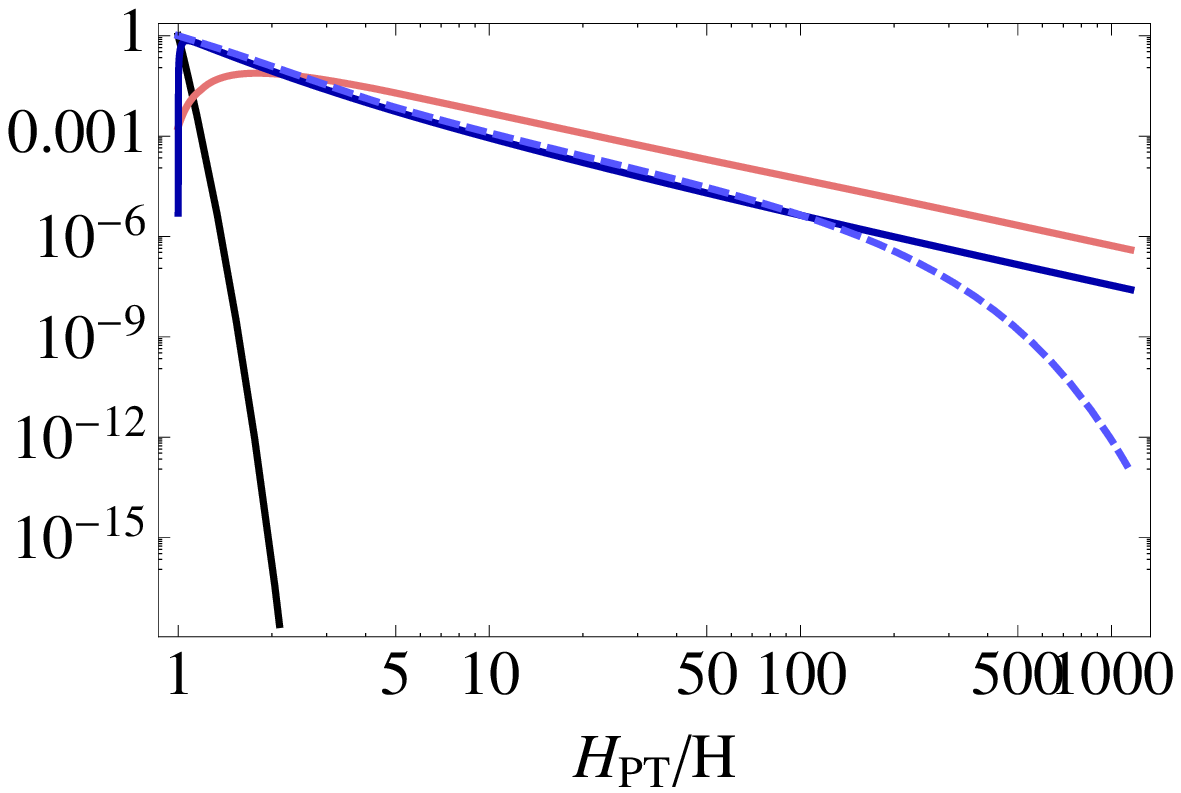}
\label{fig:left_md3}}
\subfigure[{\bf Right Panel}: $(m_\phi, f_a)=(1\, \text{TeV},10^{10}\, \text{GeV})$]{
\includegraphics[width=0.45\columnwidth,clip]{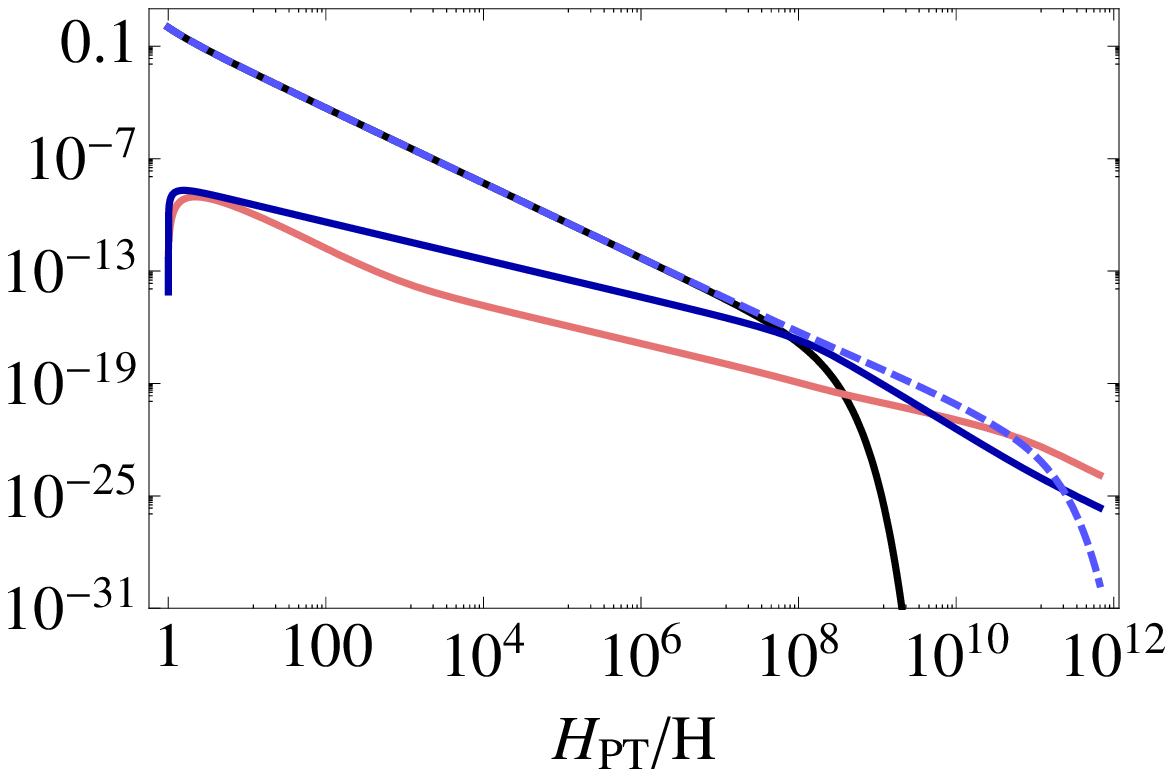}
\label{fig:right_md3}}
\caption{\small 
{\bf Left/Right Panel} shows evolution of various quantities
as a function of $H$ normalized by
one at the PQ phase transition, $H_\text{PT}$:
$\rho_\phi/\rho_{\phi, \text{PT}}$ [black],
$\rho_\text{rad}/\rho_{\phi, \text{PT}}$ [{\color[rgb]{1.000000,0.400000,0.400000}pink}],
$\rho_a/\rho_{\phi,\text{PT}}$ [{\color{blue}blue}],
$\rho_{a'}/\rho_{\phi,\text{PT}}$ [{\color{blue}blue-dashed}],
with $\rho_\bullet$ being the energy density of $\bullet$.
Note that $\rho_\phi$ and $\rho_a$ denote the energy density of the
saxion coherent oscillation and that of
the axion produced from it respectively,
and both does not contain that from thermal plasma.
See also footnote~\ref{fn:thermal}.
}
\label{fig:ev_md3}
\end{figure}

Importantly, there also exists a heavy CP-odd scalar $a'$ with a mass $\sim m_\phi$
as a mixture of the angular component of 
$\phi$ and $\bar\phi$. After the end of thermal inflation, the heavy CP-odd scalar 
may also begin an oscillation with an amplitude $\sim f_a$.
Its dissipation rate is expected to be the same as that of the axion except for the redshift factor,
\begin{align}
	\Gamma_{a'}^\text{(dis)}
	\simeq \frac{\alpha_s^2 T^3}{32 \pi^2 f_a^2}
	\times C \frac{m_\phi^2}{g_s^4 T^2} f(x),
\end{align}
where $x = m_\phi / (g_s^4 T)$.
Hence above the contour shown in Fig.~\ref{fig:cntr}
it is expected to be dissipated away into the thermal plasma.
On the other hand, below this contour 
the heavy CP-odd scalar survives from interactions with the thermal plasma,
and later it decays into gluons/gluinos through one-loop processes. 
Since the entropy production becomes more significant for a smaller mass, $m_\phi$,
the overproduction of the axion dark radiation can be avoided in all the region shown in Fig.~\ref{fig:cntr}
owing to the combination of the thermal dissipation and of the late time entropy production
from the heavy CP-odd scalar.
Fig.~\ref{fig:ev_md3} shows evolution of same quantities in Fig.~\ref{fig:ev} 
plus the energy density of heavy CP-odd scalar after the PQ phase transition
with $(m_\phi, f_a) = (50\,\text{PeV}, 10^{10}\,\text{GeV})$ and $(1\,\text{TeV}, 10^{10}\,\text{GeV})$
in the left and right panel respectively.
Here we have simply assumed that the initial energy density of heavy CP-odd scalar is 
the same as that of saxion.
As one can see from the right panel [Fig.~\ref{fig:right_md3}],
the late time decay of the heavy CP-odd scalar successfully dilutes the axion dark radiation.

\subsection{Dilution of unwanted relics and baryon asymmetry}
\label{sec:dil}

Thermal inflation dilutes the preexisting unwanted relics such as the
moduli, Polonyi and gravitinos as well as the baryon number.  
If we consider the case where the thermal dissipation is efficient,
$f_a$ is relatively small while $m_\phi$ is of the order of PeV or 
larger. Then, the $e$-folding number of the thermal inflation
is not so large.  
In the present case, the dilution factor $\Delta$ is estimated to be
\begin{equation}
  \Delta \sim \left( \frac{\sqrt{m_\phi f_a}}{T_{\rm end}} \right)^3 \sim 
  10^6 \lambda^3\left( \frac{1\,{\rm PeV}}{m_\phi} \right)^{3/2} \left( \frac{f_a}{10^{10}\,{\rm GeV}} \right)^{3/2}.
\end{equation}
For example, the final baryon asymmetry after thermal inflation is evaluated by multiplying the original baryon asymmetry
by the inverse of this factor:
\begin{equation}
	\frac{n_B}{s} = \frac{1}{\Delta}\left( \frac{n_B}{s} \right)_{\rm reh},
\end{equation}
where $(n_B/s)_{\rm reh}$ denotes the baryon-to-entropy ratio evaluated as if there were no thermal inflation.
Since the dilution factor is not so huge in the typical parameters for which the dissipative effect is efficient,
it is possible that the preexisting baryon asymmetry, created e.g. by the Affleck-Dine mechanism~\cite{Affleck:1984fy,Dine:1995kz}, 
survives the dilution due to thermal inflation and it explains the present baryon asymmetry of the universe.
Note also that the temperature after thermal inflation increases to $T_\text{max} \sim 10^{7-8}$\,GeV,
and hence it is also possible that thermal leptogenesis~\cite{Fukugita:1986hr} works if there is a mild degeneracy among right-handed neutrino masses~\cite{Pilaftsis:2003gt,Garny:2011hg}.

On the other hand, the above dilution factor is not sufficient to solve the cosmological Polonyi/moduli problem.
If there is a singlet Polonyi/moduli field, its abundance is given by
\begin{equation}
	\frac{\rho_z}{s} = \frac{1}{\Delta} \frac{T_{\rm R}^{\rm (inf)}}{8}\left( \frac{z_i}{M_P} \right)^2
	\sim 
	\frac{10^2\,{\rm GeV}}{\lambda^3}\left( \frac{m_\phi} {1\,{\rm PeV}}\right)^{3/2} \left(\frac{10^{10}\,{\rm GeV}}{f_a} \right)^{3/2}
	\left( \frac{T_{\rm R}^{(\rm inf)}} {10^9\,{\rm GeV}}\right)\left( \frac{z_i}{M_P} \right)^2,
\end{equation}
where $T_{\rm R}^{\rm (inf)}$ is the reheating temperature after
inflation and $z_i$ is the initial amplitude of the Polonyi field.  If
the Polonyi mass is of the order of the gravitino, its decay produces
too much LSPs.  Actually, in the model 2 discussed above, we may need
a singlet Polonyi field in order to make the
axino heavy.  In this case the cosmological Polonyi problem is not
solved by thermal inflation, and we need to introduce a (small)
R-parity violation to make the LSP unstable or to rely on the
adiabatic suppression mechanism~\cite{Linde:1996cx,Nakayama:2011wqa}
to suppress the Polonyi abundance.  In the model 3, on the other hand,
we do not need such a singlet Polonyi field and hence there is no
Polonyi problem.\footnote{ The Polonyi problem in the dynamical SUSY
  breaking scenario was studied in
  Ref.~\cite{Nakayama:2012hy,Evans:2013nka}.  }

\section{Discussion}  \label{sec:dis}

In this paper we have revisited the cosmology of axion models.
  If the curvature of the potential of the PQ scalar field is smaller
  than the PQ scale, like in a class of SUSY axion models, it is
likely that the PQ scalar field is trapped at the origin and it causes
thermal inflation.  We have discussed several possibilities to avoid
the overproduction of the axion dark radiation from the saxion decay
after the thermal inflation.  In general, the overproduction
constraint is relaxed by taking thermal dissipative effects on the
saxion coherent oscillation into account.  This is because, owing to
the interaction with the thermal plasma, the PQ scalar can participate
in the thermal plasma even after the PQ phase transition.  This
scenario works for the mass scale, $m_\phi$, larger than 
$\mathcal O(10)$\,TeV -- $\mathcal O(10)$\,PeV for $f_a\sim 10^9$\,GeV -- $10^{10}$\,GeV.
We have demonstrated it with some
explicit models of the saxion stabilization.  In particular, we have
shown that if there exists a heavy CP-odd scalar, depending on the
stabilization model, it can dilute the axion dark radiation due to its
late time decay.  As a result, the overproduction constraint is
further relaxed.
Those studies have significant implications to the case of high-scale inflation.
However, we note here that our results are not restricted to such a case.

One interesting prediction of the SUSY axion model is that it naturally deforms the gaugino masses~\cite{Abe:2001cg,Nakayama:2013uta}.
If the gaugino masses are dominated by the anomaly mediation effect, the additional contribution from the PQ multiplet is generally 
the same order.
Therefore, if gauginos are found and their masses are measured at the LHC, it may indicate a signal of the axion model.

Hereafter we comment on the recent claim of the discovery of the B-mode polarization in the cosmic microwave background anisotropy by the BICEP2 experiment~\cite{Ade:2014xna}, which indicates the high-scale inflation, such as the chaotic inflation~\cite{Linde:1983gd}.
The derived inflation scale in terms of the Hubble scale is as high as $H_{\rm inf} \simeq 10^{14}$\,GeV.

A prediction of the present scenario is the suppression of inflationary gravitational waves (GWs) at the high frequency range
at which future space laser interferometers have high sensitivities~\cite{Smith:2005mm}.
The energy density of inflationary GWs with frequency $f > f_c$ where
\begin{equation}
	f_c \sim 3\,{\rm Hz}\left( \frac{\sqrt{m_\phi f_a}}{10^8\,{\rm GeV}} \right),
\end{equation}
is suppressed by the factor $\Delta^{4/3}$
and hence it may be difficult to detect~\cite{Jinno:2011sw} if the dilution factor $\Delta$ is sizable.
Instead, if the PQ phase transition is first order, there is a possibility that GWs from bubble collisions
will be in the observable range~\cite{Easther:2008sx}.

So far we have focused on the hadronic axion model, because it is the simplest setup for giving the domain wall number one.
In the DFSZ axion model~\cite{Dine:1981rt,Kim:1983dt}, the PQ scalar couples to the Higgs multiplets.
However, in this model the domain wall number is larger than one, and hence there is a serious domain wall problem
if the PQ symmetry is restored during inflation.
In order to make the domain wall number one, we need to introduce five pairs of additional PQ quarks.
Thus the coupling to the PQ quarks leads to thermal inflation as studied in this paper.
An important feature in this model is that the saxion can dominantly decay into the Higgs boson pair.
Hence we do not need thermal dissipation effects to avoid the axion overproduction.

Finally we comment on the case in which the PQ symmetry is broken during inflation.
In such a case, the high scale inflation may be excluded because of too large axion isocurvature perturbation.
This argument has some loopholes.
During inflation, the saxion may have a field value much larger than $f_a$ if the saxion obtains the negative Hubble induced mass term.
Then the magnitude of axion isocurvature fluctuation is highly suppressed~\cite{Linde:1990yj}.
The fate of the saxion field with such a large initial amplitude strongly depends on the saxion stabilization model.
In the model 2, the saxion can have a field value of $\sim M_P$ during inflation,
but it is eventually trapped at the origin after the onset of coherent oscillation as extensively studied in Ref.~\cite{Moroi:2013tea}.\footnote{
	See also Refs.~\cite{Kasuya:1996ns,Kawasaki:2013iha} for the topological defect formation due to the axion self interaction.
}
The subsequent dynamics is the same as that studied in this paper.
In the model 3, the saxion tracks the temporal minimum $|\phi| \sim (HM^{n-2})^{1/(n-1)}$ 
as the Hubble parameter decreases and it relaxes to the true minimum without restoration of the PQ symmetry
if the coupling, $\lambda$, or the reheating temperature, $T_\text{R}^{\rm (inf)}$, is suppressed~\cite{Nakayama:2012zc}.
For $n=3$, for example, the magnitude of CDM isocurvature perturbation is given by
\begin{equation}
  S_{\rm CDM} = \frac{\rho_a}{\rho_{\rm CDM}} 
  \frac{H_{\rm inf}^{1/2}}{\theta_i \pi M^{1/2}}
  \sim 3\times 10^{-6}\left( \frac{\theta_i}{10^{-2}} \right)
  \left( \frac{f_a}{10^{11}\,{\rm GeV}} \right)^{1.19}
  \left( \frac{H_{\rm inf}}{10^{14}\,{\rm GeV}} \right)^{1/2}
  \left( \frac{10^{18}\,{\rm GeV}}{M} \right)^{1/2},
\end{equation}
where $\theta_i$ is the initial misalignment angle and $\rho_a$ and
$\rho_{\rm CDM}$ are the energy density of axion and CDM,
respectively.\footnote{ The misalignment angle cannot be tuned beyond
  $\theta_i \lesssim (H_{\rm inf}/M)^{1/2}$ since otherwise the
  quantum fluctuation would be dominant over the classical
  displacement of the axion field.  } Since the Planck data combined
with the WMAP9 polarization data gives an upper bound as $S_{\rm CDM}
\lesssim 1\times 10^{-5}$~\cite{Ade:2013uln},
the isocurvature constraint can be avoided.  However, the axion cannot
be the dominant component of CDM in this case.  See also
Refs.~\cite{Higaki:2014ooa,Choi:2014uaa,Chun:2014xva} for recent
discussion in this direction.  The saxion coherent oscillation with an
amplitude of $\sim f_a$ remains in this case, but it is not
cosmologically problematic~\cite{Nakayama:2012zc}.

\section*{Acknowledgments}

The work of T.M. is supported by the Japan Society for Promotion of
Science (JSPS) KAKENHI (No.~26400239 and No.~60322997).  The work of
K.N. is supported by the JSPS Grant-in-Aid for Young Scientists (B)
(No.~26800121).  The work of K.M. and M.T. are supported in part by JSPS
Research Fellowships for Young Scientists.
The work of M.T. is also supported by the program for Leading Graduate Schools, MEXT, Japan.

\appendix
\section{One-loop estimation of soft axion dissipation}
\label{sec:app}
In this appendix,
we estimate the thermal dissipation rate of soft relativistic axions with $p \ll g_s^2 T$
at the one-loop order. In principle the following estimation may be made more precise
by the resummation of many diagrams,
but we postpone this task as a future work.

The dissipation rate of relativistic axion is given by
\begin{align}
  \Gamma_a^\text{(dis)} =
  \left. \frac{\Pi_\text{J} (P)}{2 p_0}\right|_{p_0=p}.
\end{align}
where 
\begin{align}
  \Pi_\text{J}(P) = \frac{\alpha_s^2}{128\pi^2 f_a^2} 
  \int_P e^{i P \cdot x} \left< \left[ \{F^{a\mu\nu} \tilde{F}^a_{\mu\nu}\} (x) 
      \{F^{a\mu\nu} \tilde{F}^a_{\mu\nu}\} (0) \right]  \right>,
\end{align}
where $\left< \bullet \right>\equiv \tr [e^{ - \beta \hat H} \bullet]
/ \tr[e^{ - \beta \hat H}]$ stands for the canonical ensemble average.
After performing angular integrations, one finds~\cite{Salvio:2013iaa}
\begin{align}
	\Pi_\text{J} (P) =& f_\text{B}^{-1} (p_0)
	\frac{d_a}{8} \frac{\alpha_s^2}{64 \pi^5 f_a^2} \frac{1}{p}
	\int_0^\infty dk  \int_{-\infty}^\infty dk_0  
	\int_{|k-p|}^{k + p} dq\, kq f_\text{B} (k_0) f_\text{B} (p_0 - k_0) \nonumber\\
	& \Bigg\{ \left( \rho_L (K) \rho_T (Q) + \rho_T (K) \rho_L (Q) \right) 
	\left( (k + q)^2 - p^2 \right) \left( p^2 - (k-q)^2 \right) +  \nonumber \\
	&\rho_T (K) \rho_T (Q) 
	\left( \left( \frac{k_0^2}{k^2} + \frac{q_0^2}{q^2} \right) 
	\left( (k^2 - p^2 + q^2)^2 + 4 k^2 q^2 \right) \right) + 8 k_0 q_0 (k^2 + q^2 - p^2) \Bigg\},
\end{align}
where $q_0 = p_0 - k_0$, $d_a = N_c^2 -1 = 8$, $f_{\rm B}$ is the Bose-Einstein distribution function,
and $\rho_{T/L}$ denotes the spectral function for the transverse/longitudinal mode
of gauge boson in the thermal plasma.
Roughly speaking, for $p_0 = p \gg g_s^2 T$, the cut contribution of spectral function becomes important since the typical center of mass energy for scattering processes becomes
larger than thermal mass, $p T \gg g_s^2 T^2$.
And for $p_0 = p \gg g_s T$, the dominant contribution is calculated by means of 
the Hard Thermal Loop approximation in Ref.~\cite{Graf:2010tv,Salvio:2013iaa}.
On the other hand, for $p \ll g_s^2 T$, the near pole contribution of spectral function
becomes important, which is smeared by interactions in the thermal plasma.

To estimate this integral, we approximate the spectral density for the transverse mode
with the Breit-Wigner form:
\begin{align}
	\rho_T (P) = \frac{2 p_0 \Gamma_p}{[p_0^2 - \omega_p^2]^2 + [p_0 \Gamma_p]^2}; 
	~~\omega_p \equiv \sqrt{p^2 + m_\infty^2},
\end{align}
where $m_\infty$ is the asymptotic mass: for instance in the case of
non-abelian gauge theory with $N_F$ flavor, it is given by $m^2_\infty
= [C(\text{ad}) + N_F T(\text{F}) ] g_s^2 T^2 / 6$ with $C(\text{ad})$
being the adjoint Casimir and T(F) being the normalization index.
Here we estimate the width by large angle scatterings, $\Gamma_p \sim
g_s^4 T^3 / p^2$~\cite{Moore:2008ws}.  Since the loop-integral is
typically dominated by a hard momentum $\sim T$, we omit contributions
from longitudinal mode $\rho_L$, which is suppressed by the
residue~\cite{text}.  After some calculations, one can express the
dissipation rate with
\begin{align}
	\Gamma_a^\text{(dis)} = \frac{ \alpha_s^2 T^3}{32 \pi^2 f_a^2}
	\times
	C\frac{p_0^2}{g_s^4 T^2} f(x),
\end{align}
where $x =p_0/g_s^4T$, and we found that the function $f$
can be approximated as
\begin{align}
\label{eq:function}
	f(x) \simeq  \frac{d_a}{\pi^2} & \int_{a} dy \frac{e^y}{\left( e^y - 1 \right)^2}  \frac{\left( y^2 - a^2 \right)^{3/2}}{y} \nonumber\\
	\frac{1}{2 b} & \int_{-b}^b   \frac{d \epsilon}{1 + g_s^4 \epsilon \sqrt{1 - \frac{a^2}{y^2}}  \frac{x}{y}}\,
	\frac{ y^2 - a^2 + y^2 \epsilon^2 + 2 \sqrt{y^2 - a^2} y \epsilon }
	{x^2 \left(1 + \sqrt{1 - \frac{a^2}{y^2}} \epsilon \right)^2 (y^2 - a^2)^2 + 1},
\end{align}
with
\begin{align}
	a &= m_\infty/T,~~
	b =
	\begin{cases}
	1 &\text{for~relativistic~axion}\\
	0 &\text{for~non-relativistic~axion}
	\end{cases}.
\end{align}
The asymptotic behavior of $f(x)$ is $f(x) \sim {\rm const.}$ for $x \ll 1$ and $f(x) \propto x^{-2}$ for $x \gg 1$.
Notice that our estimation of $\Gamma_a^\text{(dis)}$ is
  expected to be an order-of-estimate of the dissipation rate because
  it is based on a simple one-loop calculation.  Thus, in order to take
  account of the uncertainty in our result, we introduce the
  numerical coefficient $C$ and vary it in our numerical analysis.


\end{document}